# Near real-time monitoring of global land-ocean cover dynamics

Lixing Wang[1], Tao Li[1], Xinyu Dou[2], and Zhu Liu[1]

[1]Department of Earth System Science, Tsinghua University, Beijing, 100084, China
[2]Doerr School of Sustainability, Department of Earth System Science, Stanford University, Stanford, 94305, US

*Correspondence to*: Xinyu Dou (douxy24@stanford.edu) and Zhu Liu (zhuliu@tsinghua.edu.cn)

**Abstract.** Monitoring the dynamics of global land-ocean cover is fundamental for regulating the Earth's climate and sustaining terrestrial and marine ecosystems. However, existing datasets and research often exhibit limitations in temporal resolution and timeliness, lack coupled analysis of land cover and sea ice dynamics, and fail to incorporate the perspective of Earth system safety thresholds. Here, we developed an integrated monitoring framework by fusing multi-source remote sensing and reanalysis data, generating a 5-day resolution time series (2018-2025) of global land cover and sea ice coverage with near-real-time update capability. Our analysis reveals distinct latitudinal and regional patterns, with forests dominating (27.0% of global land area) tropical and subtropical regions. At the national scale, land cover composition and seasonal rhythms vary significantly, with countries like China, India, and the US exhibiting divergent patterns such as bimodal cropland fluctuations and alternating snow/ice dominance. Temporally, vegetated cover types exhibit seasonal cycles peaking during Northern Hemisphere summer, and a pronounced anti-phase seasonal pattern is observed between Arctic and Antarctic sea ice coverage. Crucially, safety threshold analysis indicates the global forest cover indicator (~60%) is approaching the 54% lower safe limit, with a declining trend in recent years. Concurrently, Arctic sea ice coverage in September occasionally drops to 23%, below its critical upper limit of 27.6%. Temperature presents a significant negative correlation with sea ice cover ($R$ = -0.78, $p$ < 0.001), with asymmetric freezing and melting rates. By quantifying the proximity of key indicators to their safety thresholds, this study provides a robust, integrated framework for early-warning assessment, thereby offering vital scientific support for global climate adaptation and sustainable policymaking.

## 1 Introduction

Global land-ocean cover, which encompasses both land cover and sea ice cover, constitutes a pivotal component of the Earth system (Turner et al., 2007; Notz and Bitz, 2017). Its dynamic changes directly influence climatic regulation and carbon cycle processes, while also profoundly affecting biodiversity distribution and the human living environment (Song et al., 2018). High-quality, long-term, and near real-time monitoring data of these dynamics are therefore fundamental for Earth system modeling and climate change attribution (Rounsevell et al., 2014). In the current context of intensifying extreme climate events and the frequent occurrence of phenomena such as global warming, sea ice ablation, and forest degradation (Shahi et al., 2023; Lapola et al., 2023), the need for accurate and timely monitoring of land-ocean cover dynamics has become increasingly urgent.

A variety of datasets based on remote sensing observation and model are currently available for the global monitoring of land-ocean cover. For land use and land cover (LULC) monitoring, mainstream datasets include the MODIS land cover product (MCD12Q1) (Friedl and Sulla-Menashe, 2022), which provides annual categorical LULC data at a spatial resolution of 500 m. The European Space Agency Climate Change Initiative (ESA-CCI) land cover product, spanning from 1992 to 2022 with a 300-m resolution, allows for a fine classification of vegetation types (Esa, 2017; Wang et al., 2023). The FROM-GLC dataset offers global land cover distribution at 30-m resolution with a classification accuracy of 74.3% (Yu et al., 2022; Yu et al., 2025), serving as a key dataset for surface monitoring and terrestrial model simulations. The Dynamic World dataset provides near real-time updated data at 10-m resolution, significantly improving the temporal resolution for capturing land cover dynamics (Brown et al., 2022). Regarding sea ice cover monitoring, several well-established sea ice concentration products have been released by institutions such as the University of Bremen, the National Snow and Ice Data Center (NSIDC), the University of Hamburg, and the European Organisation for the Exploitation of Meteorological Satellites (EUMETSAT), with spatial resolutions ranging from 25 km to 6.25 km (Spreen et al., 2008; Windnagel et al., 2021; Kaleschke et al., 2016; Tonboe et al., 2016). The ERA5 reanalysis product from the European Centre for Medium-Range Weather Forecasts (ECMWF) provides sea ice cover data that support hourly near real-time updates since 1940 (Hersbach et al., 2020). By leveraging these datasets, existing studies have preliminarily revealed the evolutionary patterns of land and sea ice cover in different regions, including key processes such as tropical forest degradation and polar sea ice ablation (Tarazona and Miyasiro-López, 2020; Macdermid et al., 2025).

Nevertheless, limitations persist in the existing research. First, a majority of the current land-ocean cover datasets suffer from a significant update lag, typically exceeding one year, which makes them unable to meet the practical demands for timely response and rapid assessment in the context of extreme climate events (Li et al., 2023; Tschudi et al., 2020). Although a few advanced datasets (e.g., Dynamic World and ERA5) enable near real-time updates of raster imagery, they still lack systematic time-series analyses of the areal changes in various global cover types, failing to fully capture the dynamic evolutionary processes of land-ocean cover (Brown et al., 2022; Hersbach et al., 2020). Second, most studies conduct monitoring of land cover and sea ice cover as separate and independent subjects, lacking a systematic coupled analysis of the two at the global scale (Liu et al., 2020). This deficiency hinders the complete characterization of the dynamic cover features under the influence of land-ocean interactions and restricts the understanding of the overall evolutionary laws of the global land-ocean cover system.

A further limitation is the absence of a perspective centered on the Earth system safety thresholds. The Planetary Boundaries theory posits that the Earth system encompasses critical threshold values for key environmental indicators, which are essential to maintaining the system's stability and safeguarding human survival and development (Rockström et al., 2009; Steffen et al., 2015). Among these indicators, Land System Change acts as a core element in regulating the global climate and preserving ecological balance (Tobian et al., 2024). Against the backdrop of the growing intensification of extreme climate events, current land-ocean cover monitoring studies have not integrated the Planetary Boundaries theory into their research frameworks, nor do they provide quantitative analyses to determine whether key cover types such as forests and sea ice have crossed the safety boundaries of the Earth system. Consequently, such research is unable to effectively underpin global-scale safety assessment and early warning of the Earth system.

In this study, we developed a near real-time and coupled framework for the dynamic monitoring of global land-ocean cover. By integrating multi-source remote-sensing and reanalysis products, we generate a 5-day resolution time series (2018–2025) of global land cover and sea ice coverage that supports near real-time updates with a lag of less than one month. The dataset enables systematic analyses of global-scale spatiotemporal dynamics, national-scale temporal changes and safety threshold assessment. This work thus provides high-timeliness data support and policy-relevant insights for global climate change response, ecological safety evaluation, and sustainable development decision-making.

## 2  Materials and methods

Based on multi-source remote sensing and reanalysis data (Table 1), this study established a comprehensive framework for near real-time data integration, coupled analysis and threshold assessment using the Google Earth Engine (GEE) platform and the Colab Python coding environment. This framework was applied to sequentially complete the preprocessing and long time series generation of global land cover data, as well as the aggregation and downscaling of global sea ice cover data. Quantitative analysis of safety thresholds was subsequently carried out for two key components, namely forest cover and sea ice cover, while the impacts of temperature variations on sea ice coverage were also investigated. Ultimately, this framework enabled the dynamic monitoring of land-ocean cover at global and national scales and the safety assessment of the Earth system (Fig. 1).

**Table 1**. Data information.

| Dataset | Variable | Spatial resolution | Temporal resolution | Source |
|---|---|---|---|---|
| Dynamic World | 9 types of LULC | 10m | 2-5 days | https://www.dynamicworld.app/ |
| ERA5 Hourly | Sea ice cover<br>2m air temperature | 0.25° | Hourly | https://cds.climate.copernicus.eu/datasets/reanalysis-era5-single-levels?tab=overview |
| Potential Forest Coverage | Potential forest cover | 0.01° | — | https://www.arcgis.com/home/item.html?id=120dce192e754c8084f61eee6a2d9edf |

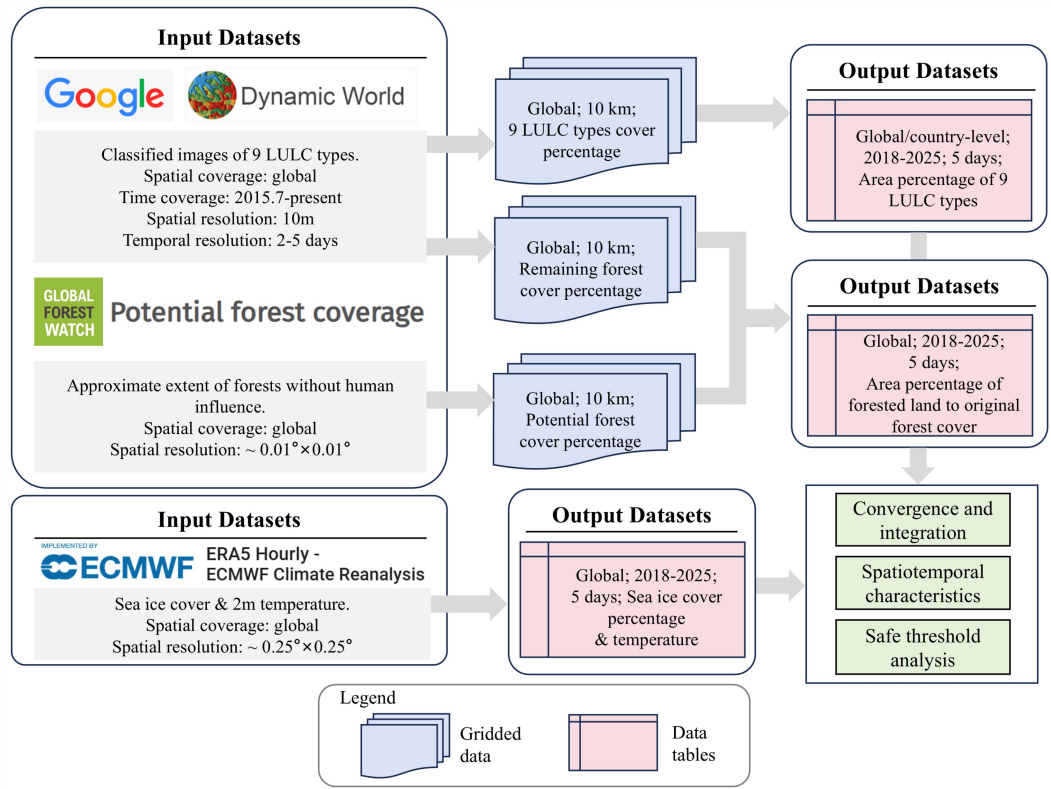

**Figure 1.** Workflow of this study.

## 2.1 Data

### 2.1.1 Dynamic World dataset

The Dynamic World dataset is built upon Sentinel-2 satellite imagery and employs deep learning models to achieve fine-grained classification of global LULC (Brown et al., 2022). It comprises nine major land cover types including forest, grassland, bare ground, built-up areas, etc., with an overall classification accuracy of 73.8%, which allows for the effective capture of dynamic variation characteristics of global land cover. The image dataset spans from July 2015 to the present, with an update frequency of 2 to 5 days depending on latitude. Its original spatial resolution is 10 m, and for the purpose of subsequent analyses in this study, the dataset is resampled to a 10 km resolution by comprehensively considering the computational costs and analytical requirements of global-scale research.

### 2.1.2 ERA5 hourly sea ice cover data

The ERA5 hourly sea ice cover data is a global reanalysis product released by ECMWF (Hersbach et al., 2020). This dataset is primarily used to characterize the spatial distribution and concentration of global sea ice, with its spatiotemporal scope spanning from 1940 to the present. It features an original spatial resolution of 0.25°, where each grid cell value represents the proportion of sea ice cover within that cell. The data quality has been rigorously validated through observational data assimilation, and its accuracy meets the requirements for global-scale dynamic monitoring of sea ice. In this study, hourly data from 2018 to 2025 were selected for processing; the dataset was resampled to a 10 km spatial resolution via area-weighted averaging and downscaling to match the resolution required for subsequent analyses.

### 2.1.3 Potential forest coverage data

The Potential Forest Coverage dataset is retrieved from the Global Forest Watch platform (Gfw, 2019). Based on key natural environmental factors including climate, soil and topography, and integrated with historical vegetation distribution records, this dataset simulates the potential distribution of global primary forests in the pre-industrial era, a period largely unaffected by significant human activities. It serves as the benchmark data for quantifying Land System Changes and assessing the safety thresholds of forest cover in this study. Characterized as a global static dataset with a spatial resolution of approximately 0.01°, it depicts the approximate extent of forests that would exist without human interference, thereby providing a fundamental basis for calculating the ratio of extant forest cover to primary forest cover.

### 2.1.4 ERA5 hourly temperature data

The ERA5 hourly temperature data is also a component of the ERA5 reanalysis product, and it is used to analyze the impacts of temperature variations on sea ice coverage in this study. Its core variable is the 2-m surface air temperature, with a spatiotemporal scope extending from 1940 to the present and an original spatial resolution of 0.25°. Calibrated through the assimilation of observational data from global meteorological stations, the dataset features high spatiotemporal continuity and accuracy. Hourly temperature data from 2018 to 2025 were processed synchronously with the sea ice cover data. Area-weighted averaging was applied to achieve spatiotemporal matching between the two dataset.

## 2.2 Methodology

### 2.2.1 Global land cover data processing

This study developed a data processing system based on the GEE platform and the Colab Python coding environment. The Dynamic World dataset was first processed to generate a long time series of the percentage changes in the area of various global land cover types for the period 2018–2025, with a temporal resolution of 5 days. For each temporal snapshot, the corresponding raster map was extracted and resampled to a 10 km spatial resolution. During resampling, the value of each 10 km grid cell was assigned as the mode of land cover classes within the cell, which effectively represents the dominant land cover characteristics in the target region. To mitigate the area statistical errors caused by partial pixel missing in the original classified imagery, a forward pixel-backfilling algorithm was applied: the missing pixel values to be processed were filled using the latest non-null pixel values from the nearest available date, with a typical backfilling time window of 1–2 weeks and a maximum of 1 month. After processing with this method, the overall data missing rate was reduced to 6.5%, which effectively ensured data integrity. In addition, spatial extraction was conducted based on national boundaries to obtain the land cover distribution maps and long time series data for each country, providing data support for national-scale analyses.

**2.2.2    Global sea ice cover data processing and analysis**

ERA5 hourly sea ice cover data were aggregated using temporal averaging and area-weighted averaging methods to generate a long time series of global sea ice coverage with a 5-day resolution for the period 2018–2025. To achieve spatial matching with the land cover data, a bilinear interpolation method was employed to downscale the original sea ice data from a 0.25° spatial resolution to 10 km, and corresponding sea ice cover maps were generated accordingly. Considering the uniqueness and importance of sea ice in the Arctic and Antarctic, the study area was zoned by the Arctic and Antarctic Circles, and in-depth analyses were conducted on the dynamic changes in sea ice cover for the two polar regions separately, so as to accurately capture the evolutionary characteristics of polar sea ice.

**2.2.3    Safety threshold analysis for key indicators**

Combined with the Planetary Boundaries theory, a quantitative analysis of safety thresholds was first conducted for global forest cover, with the Land System Change indicator of the Planetary Boundaries as the key variable (Steffen et al., 2015). Based on the GEE platform, the global spatial distribution map of forest cover was extracted from the Dynamic World dataset, which was used as the distribution data for the current remaining forest. The Potential Forest Coverage data was adopted as the baseline for the analysis. Taking each 10 km grid cell as the basic unit, the area proportions of forest cover in each grid cell were calculated separately for the two datasets. The ratio of these two proportions was then derived to obtain the control variable of the Land System Change indicator, namely the area of forested land as the percentage of original forest cover, which was calculated as follows:

$$Pct_i = \frac{Pct_{i,\ remaining}}{Pct_{i,\ original}} \times 100\% \tag{1}$$

where $i$ denotes the index of each 10 km grid cell, $Pct_{i,remaining}$ represents the percentage of currently remaining forest within the grid cell, $Pct_{i,original}$ denotes the percentage of original forest in the grid cell, and $Pct_i$ is the derived control variable for the Land System Change indicator. The safety threshold range for this control variable is set at 54% to 75% (Steffen et al., 2015), which serves as the criterion for determining whether the current state of forest cover falls within the safe boundaries of the Earth system.

For the safety threshold of sea ice cover, this study focuses on the sea ice area coverage in the Arctic during September. With reference to the cryosphere threshold report by the International Cryosphere Climate Initiative and the IPCC Sixth Assessment Report (AR6) (Icci, 2015; Arias et al., 2021), an Arctic sea ice extent of ⩾ 4 million km² in September is regarded as the critical threshold for maintaining Arctic climatic stability, while a sea ice extent of <1 million km² is identified as an irreversible climatic tipping point corresponding to a nearly ice-free Arctic. The above areal thresholds were converted into the percentage of the Arctic total area (approximately 6.9%–27.6%), which was matched with the sea ice coverage data generated in this study and used as the criterion for the sea ice cover safety threshold. Meanwhile, we conducted an analysis of the impacts of temperature variations on changes in sea ice coverage. The ERA5 temperature dataset was processed via temporal averaging and area-weighted averaging, then matched with the sea ice cover data for correlation analysis. The Pearson correlation coefficient ($R$) and $p$-value were used to evaluate the strength of the correlation between the two variables.

## 3 Results

### 3.1 Spatiotemporal dynamics of global land-ocean cover

The spatial distribution pattern of global land-ocean cover (Fig. 2a) clearly exhibits the latitudinal gradients and regional differentiation among cover types. Forests constitute the largest proportion of global vegetated cover, occupying approximately 27.0% of the global land area. They are predominantly concentrated in tropical and subtropical regions such as the Amazon Basin, the Congo Basin, and Southeast Asia, forming the world's core carbon sinks and biodiversity hotspots (Mehta et al., 2025; Worden et al., 2026). Shrubland and grassland are widely distributed in inland and semi-arid areas of the mid-low latitudes, accounting for 12.9% and 4.9% of the global land area respectively, and constitute transitional ecological zones. Cropland is concentrated in the mid-latitude agricultural regions of the Northern Hemisphere (e.g. Eastern Europe and South Asia) with a proportion of 8.3%, reflecting the spatial agglomeration of human agricultural activities.

Among non-vegetated cover types, built-up areas constitute a mere 1.5% of the global land area, yet are highly concentrated in coastal urban agglomerations and major economic cores. Snow and ice are extensively distributed near the Arctic and Antarctic Circles, making up 22.1% of the global land and marine marginal areas, while bare ground accounts for 21.7%, forming the basic surface of arid regions. The remaining areas are covered by water bodies, with inland water accounting for 1.3% of the global land area. Notably, sea ice coverage in the Arctic exceeds 85%, and coverage near both polar circles decreases with declining latitude and increasing distance from continental margins. The coupling of sea ice cover data and land cover data compensates for the limitation of a single classification for marine water bodies in traditional land cover classification systems, and effectively distinguishes the complex patterns of sea ice coverage in coastal and polar regions.

Figure 2b reveals the seasonal characteristics of the area proportions of various global land-ocean cover types from 2022 to 2025, with distinctly different fluctuation patterns observed between vegetated types (left) and non-vegetated types (right). Among vegetated types, forest, as the dominant type, has an area proportion stably ranging from 17% to 40% with a large interannual fluctuation amplitude, which is mainly driven by seasonal disturbances of tropical forests and regional deforestation/afforestation activities (De Sales et al., 2020). Cropland and shrubland maintain their proportions in the ranges of 7%–11% and 10%–17% respectively, exhibiting synchronous seasonal fluctuations that reflect the coupling effect of agricultural production rhythms and the phenology of natural vegetation (Liu et al., 2017). Although grassland and flooded vegetation account for relatively low proportions, they show distinct seasonal patterns similar to other vegetated types. Dominated by the extensive distribution of vegetation in the Northern Hemisphere, the total proportion of vegetated types peaks from June to September each year, which corresponds to the vigorous growth of vegetation in summer. In contrast, among non-vegetated types, land snow and ice cover displays an entirely opposite seasonal trend, hitting a trough from June to September with its area proportion dropping from 40% to 7%. The remaining non-vegetated types show stable fluctuations with only slight seasonal variations.

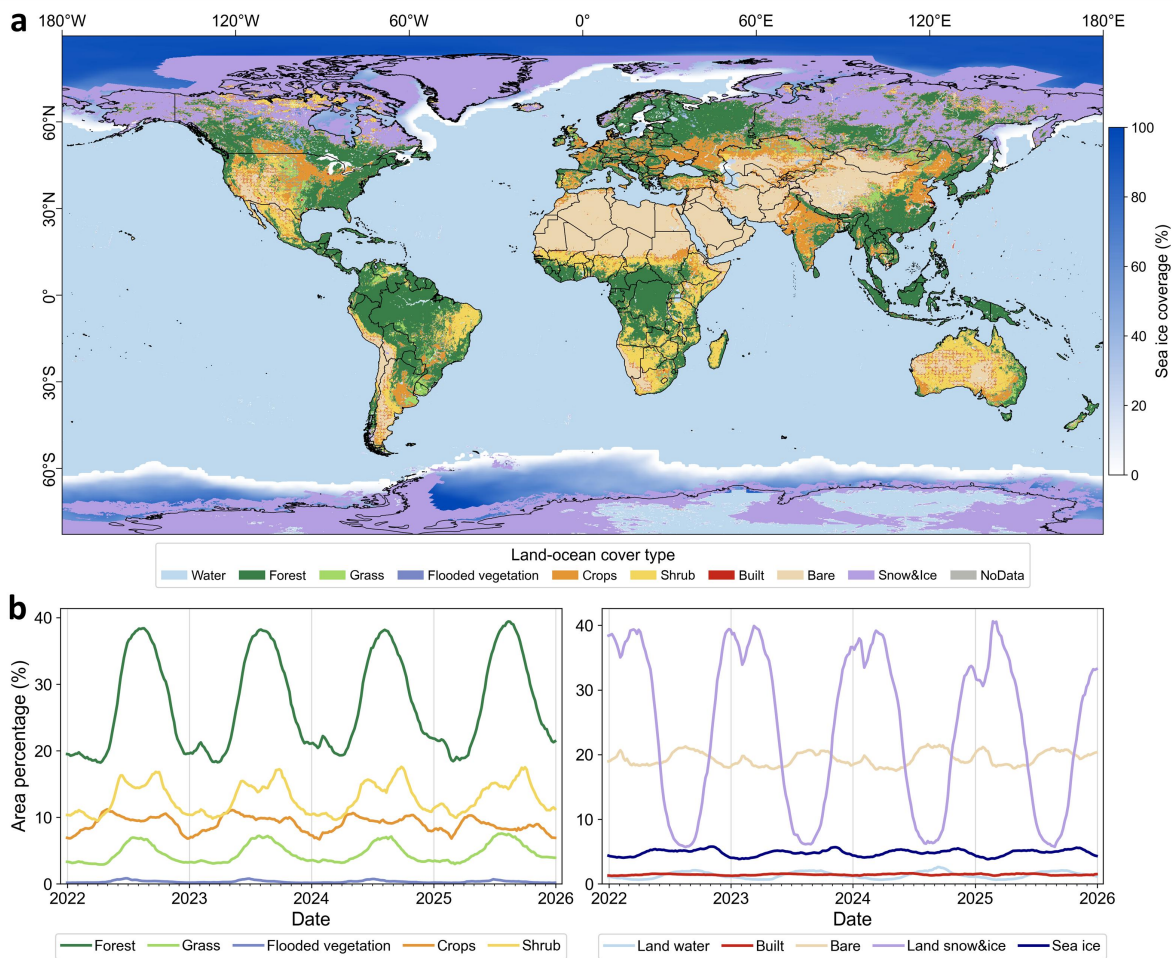

**Figure 2. a,** The spatial distribution of global land-ocean cover in 2025. **b,** The temporal change of the area percentage of each type from 2022 to 2025. The percentage of sea ice cover is calculated by the global ocean area, and the other nine types are calculated by the global land area. The temporal change from 2018 to 2021 is shown in Fig. S1.

Table 2 further illustrates the long-term annual trends of global land-ocean cover from 2018 to 2025. Forest, flooded vegetation, and crops show continuous declining trends in the past three years (2023-2025): forest decreased from 27.57% to 27.04%, flooded vegetation dropped from 0.35% to 0.31%, and crops declined notably from 9.39% to 8.32%. In contrast, built-up areas and bare land exhibited overall increasing trends from 2018 to 2025, with built-up areas rising from 1.31% to 1.47% and bare land expanding from 19.62% to 21.66%, implying that ecological land types have been converted to human-dominated and bare surfaces. Land snow & ice also decreased from 24.01% (2022) to 22.08% (2025), consistent with cryosphere degradation under global warming. Grassland, shrubland, land water and sea ice remained relatively stable with minor interannual fluctuations. These long-term trends supplement the seasonal dynamics, collectively characterizing both periodic and secular changes in global land-ocean cover and providing intuitive temporal evidence for global ecological and environmental shifts.

**Table 2**. Annual area percentage of global land-ocean cover from 2018 to 2025 (unit: %).

| Year | Land water | Forest | Grass | Flooded vegetation | Crops | Shrub | Built | Bare | Land snow&ice | Sea ice |
|---|---|---|---|---|---|---|---|---|---|---|
| 2018 | 1.297 | 26.846 | 4.670 | 0.379 | 9.320 | 12.922 | 1.314 | 19.621 | 23.632 | 4.907 |
| 2019 | 1.344 | 26.871 | 4.901 | 0.381 | 9.634 | 12.917 | 1.353 | 18.432 | 24.167 | 4.888 |
| 2020 | 1.421 | 27.420 | 4.714 | 0.375 | 9.599 | 12.759 | 1.384 | 19.023 | 23.306 | 4.985 |
| 2021 | 1.380 | 26.998 | 4.507 | 0.374 | 9.520 | 12.929 | 1.395 | 19.004 | 23.894 | 5.076 |
| 2022 | 1.338 | 27.002 | 4.446 | 0.383 | 9.350 | 12.897 | 1.429 | 19.146 | 24.008 | 4.904 |
| 2023 | 1.375 | 27.573 | 4.746 | 0.352 | 9.390 | 12.995 | 1.456 | 18.450 | 23.664 | 4.735 |
| 2024 | 1.385 | 27.164 | 4.479 | 0.346 | 9.073 | 12.877 | 1.463 | 20.120 | 23.093 | 4.843 |
| 2025 | 1.345 | 27.044 | 4.888 | 0.314 | 8.319 | 12.890 | 1.469 | 21.655 | 22.076 | 4.833 |

## 3.2 Temporal characteristics of national-scale land cover and polar sea ice cover

Figure 3a–f and Fig. S2 illustrate the temporal characteristics of the proportions of land cover types in 12 major global countries from 2022 to 2025. Owing to disparities in climate, agricultural practices and the intensity of human activities, distinct dominance of specific cover types and seasonal rhythms are observed across different countries. In China, forest, bare ground and cropland are the dominant land cover types, collectively accounting for over 70% of the total country area. Among these, the proportion of cropland exhibits an obvious bimodal fluctuation pattern, with the primary peak in May and the secondary peak in November each year. This pattern arises from the superposition of multiple cropping systems across northern and southern China and the phenological rhythms of crops (Zhou et al., 2025; Xiao et al., 2025). In India, forest and cropland dominate the land cover, with the proportion of cropland stably ranging from 25% to 45% and showing a smaller amplitude of seasonal fluctuation than that in China, which embodies the characteristics of continuous agricultural utilization under the multiple cropping system in tropical monsoon regions (Wang et al., 2022).

The land cover pattern of the United States features a unique dual-dominance of forest and snow&ice, which alternately account for a high proportion of the land area. The proportion of snow&ice rises sharply in winter (December–February) and recedes rapidly in spring (March–May), which is highly consistent with the seasonal snow melting process under the temperate continental climate of North America. For European countries (the UK, France, Germany, etc.), forest, cropland and grassland serve as the core land cover types. The proportion of forest can reach over 40% at its peak, the proportion of cropland hits a trough at the end of the year, and the proportion of grassland rises slightly in winter. These features reflect the vegetation senescence and flourishing patterns in the transitional zone between the temperate maritime and continental climates. Overall, the temporal variations in national-scale land cover all exhibit distinct annual cycles, yet the fluctuation amplitudes vary significantly across different types: the bimodal fluctuation of cropland is the most pronounced in agriculture-dominated countries (China and India); the fluctuation of snow & ice is the most striking in climate-sensitive countries (the United States and the UK); while forest displays a stable unimodal fluctuation in all countries, acting as the fundamental cover type of terrestrial ecosystems.

Figure 3g reveals the inverse seasonal patterns of sea ice cover proportions in the Arctic and Antarctic from 2022 to 2025. Arctic sea ice exhibits a clear unimodal annual cycle: it reaches a peak of approximately 80% between February and April and drops to a trough of about 20%–25% during September and October, with a stable fluctuation amplitude of 50 percentage points, which embodies the seasonal freezing and thawing dynamics driven by the temperate-frigid climate of the Northern Hemisphere. Antarctic sea ice shows a unimodal cycle with an opposite phase: it peaks at around 90% from September to October and falls to a trough of about 15% during February and March, with a larger fluctuation amplitude of approximately 75 percentage points, reflecting the seasonal asymmetry of the polar climate in the Southern Hemisphere. Such asymmetry in the variations of Arctic and Antarctic sea ice is not only associated with the differences in climatic systems between the two hemispheres but also reflects the divergent responses of polar ecosystems against the backdrop of global warming.

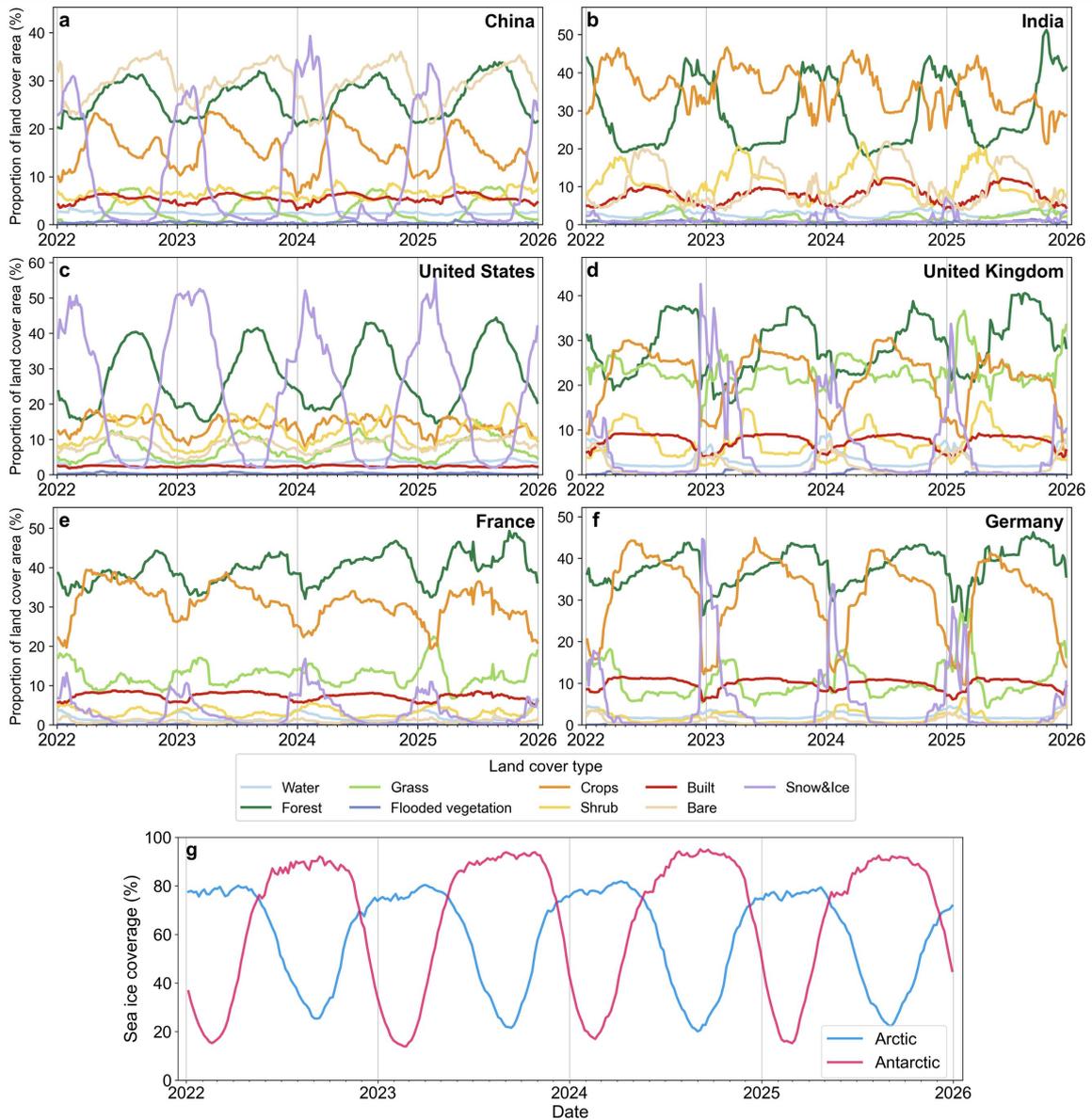

**Figure 3. a-f,** Temporal changes of land cover in 6 major countries (China, India, the United States, the United Kingdom, France and Germany). Temporal changes of the other 6 countries are shown in Fig. S2. **g,** Temporal changes of sea ice coverage of the Arctic and Antarctic.

## 3.3 Quarterly fluctuations of land-ocean indicators near safe limits

We further analyzed the seasonal deviations of global forest coverage and sea ice coverage from their respective safe thresholds. Figure 4a–d depicts the quarterly spatiotemporal distribution of global forest and sea ice coverage, revealing the spatiotemporal evolution patterns of these two indicators. The area percentage of extant forested land relative to original forest cover serves as the control variable for Land System Change in the Planetary Boundaries theory, which is used to reflect the health status of the global natural ecosystem (Steffen et al., 2015). Spatially, this indicator exhibits prominent latitudinal differentiation and regional agglomeration characteristics: high-value zones (>70%) are mainly concentrated in tropical primary forest distribution areas such as the Amazon Basin, the Congo Basin and the Southeast Asian archipelagos, where forest coverage is close to the level of primary forests; most temperate forest regions in the mid-to-high latitudes (e.g., North America and the Russian Far East) show values below 60%. In contrast, the indicator values are significantly smaller than the lower limit of the safe threshold (54%) in the vicinity of densely populated urban agglomerations and some agricultural development zones, forming distinct high-risk zones of forest degradation. Temporally, the spatial patterns vary markedly across quarters, with opposing peak periods observed between the Northern and Southern Hemispheres. For instance, the forest coverage in Russia reaches its peak in the third quarter, during which the Amazon rainforest hits its trough, while the amplitude of seasonal variation in the Southern Hemisphere is less pronounced than that in the Northern Hemisphere.

Figure 4e quantitatively presents the time series of the global forest cover indicator from 2018 to 2025 and its relationship with the safe limit range (54%–75%). The results show that the global mean value of the forest cover indicator remains stably within the safe limit at approximately 60%, slightly above the lower limit, indicating that the global forest system is approaching the safe boundary of the Earth system (Kitzmann et al., 2025). The time series curve displays distinct seasonal fluctuations: forest coverage exceeds the upper limit during peak periods (typically the growing season) and falls below the lower limit during trough periods. Notably, in early 2020, affected by the COVID-19 pandemic, the trough value was significantly higher than in other years, reflecting the reduced pressure on forest land use due to diminished human activities under large-scale lockdowns (Chen and Yang, 2021; Houballah et al., 2024). Table 3 summarizes the annual mean values of the forest cover indicator, which exhibit a declining trend over the past three years (2023-2025), warning that global forest degradation is trending toward breaching the lower safe limit.

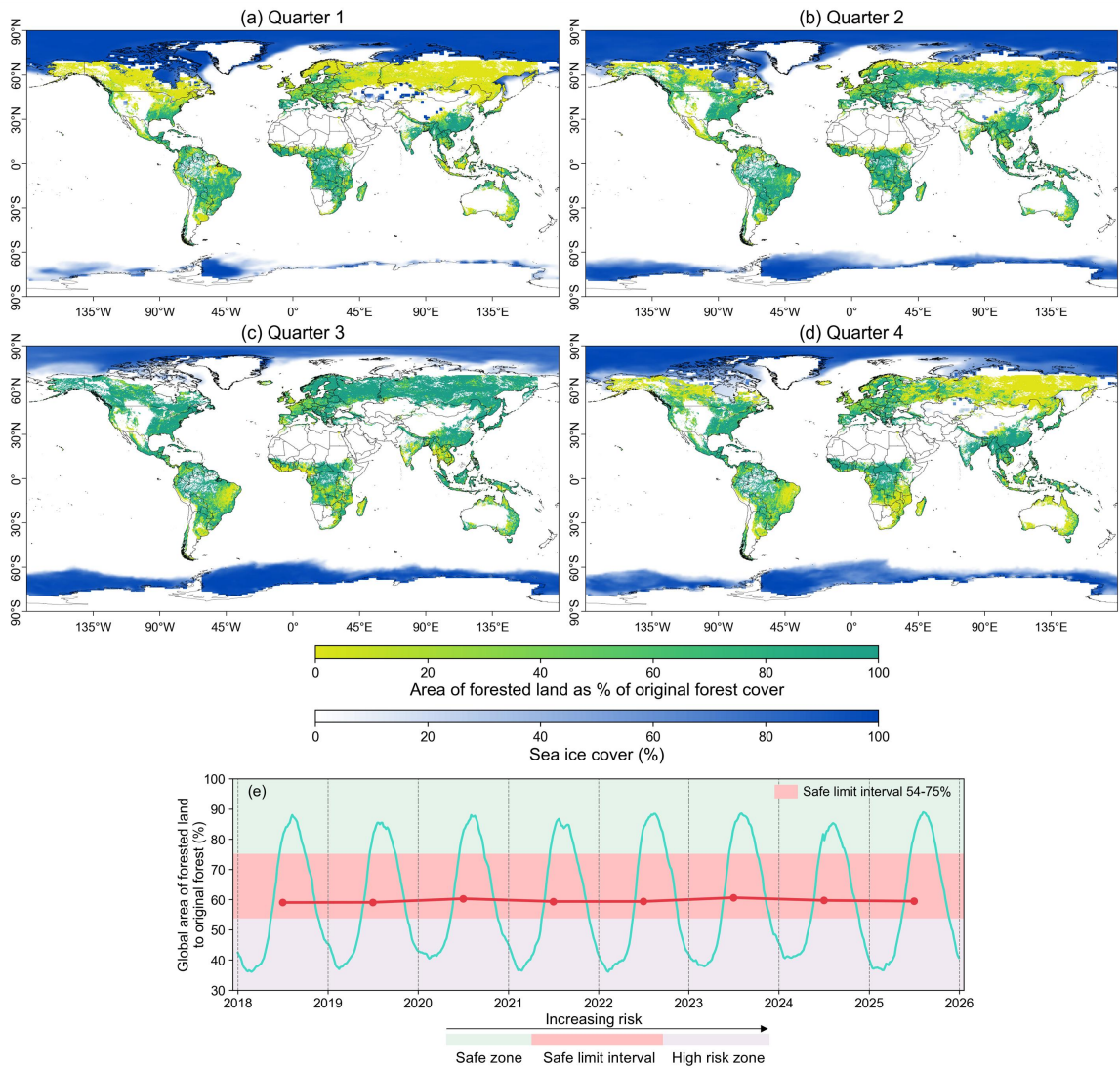

**Figure 4. a-d,** Quarterly distribution of global forest and sea ice cover indicators. **e,** Temporal changes of global area of forested land to original forest. The red line represents the annual average.

**Table 3**. Annual average of forest and sea ice cover indicators (unit: %).

| Year | Global area of forested land to original forest | Arctic sea ice coverage | Antarctic sea ice coverage |
| --- | --- | --- | --- |
| 2018 | 59.04 | 58.87 | 66.20 |
| 2019 | 59.10 | 58.20 | 65.84 |
| 2020 | 60.30 | 57.53 | 66.92 |
| 2021 | 59.38 | 60.53 | 68.95 |
| 2022 | 59.38 | 60.76 | 63.46 |
| 2023 | 60.64 | 59.88 | 66.53 |
| 2024 | 59.74 | 59.17 | 67.61 |
| 2025 | 59.48 | 58.19 | 65.81 |

Regarding the quarterly variations in sea ice coverage, opposite trends are also observed between the Arctic and Antarctic. Fig. 4a–d shows that Arctic sea ice coverage peaks in Quarter 1 and retreats in Quarter 3, whereas Antarctic sea ice expands continuously from Quarter 1 to Quarter 3. The IPCC AR6 report identifies September Arctic sea ice extent as a key indicator for assessing the safety of sea ice conditions (Arias et al., 2021). Figure 5a illustrates the long-term variation in Arctic sea ice coverage, revealing that Arctic sea ice coverage reaches its trough (~23%) in September and slightly drops below the upper bound of the safe limit (Noaa, 2024; Amap, 2025). However, it remains above the safe limit for most of the year, with an annual average of approximately 60%.

The seasonal trend of sea ice coverage is mainly influenced by temperature. As illustrated in Fig. 5b, the scatter plot of Arctic sea ice coverage versus corresponding mean temperature reveals a significant overall negative correlation ($R = -0.78$, $p < 0.001$). Notably, data points corresponding to the periods from winter to summer and from summer to winter form two separate groups, indicating that the rates of sea ice increase and decrease are asymmetric. Specifically, the melting rate during warming phases is slower than the freezing rate during cooling phases. This asymmetry aligns with principles of thermodynamics and phase transition dynamics (Maykut and Untersteiner, 1971; Untersteiner, 2013): the ice-to-water transition, driven by rising temperatures, requires overcoming intermolecular hydrogen bonds and the latent heat barrier, with energy input efficiency and interfacial heat transfer jointly limiting the melting rate. In contrast, the water-to-ice transition during cooling is facilitated by ice crystal nucleation and growth, which more readily satisfies the condition of decreasing Gibbs free energy; the continuous release of latent heat during freezing further accelerates ice expansion, making freezing thermodynamically more favorable than melting. The same asymmetric pattern applies to changes in Antarctic sea ice coverage (Fig. S3). With the frequent occurrence of extreme high-temperature events, the impact of temperature on sea ice melting has become increasingly profound. Since 2022, Arctic sea ice coverage has shown a declining trend (Table 3), reflecting the urgency of maintaining sea ice coverage under global warming.

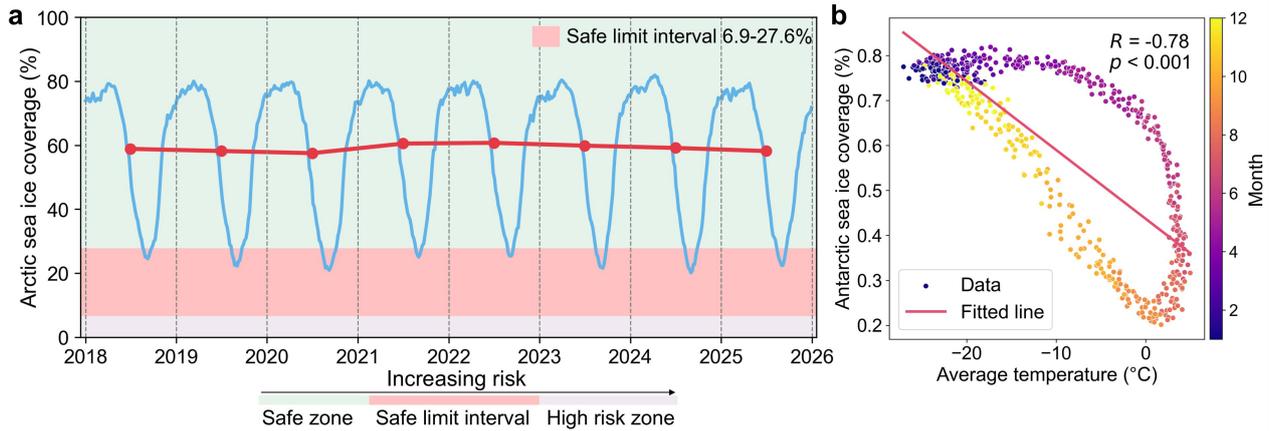

**Figure 5. a,** Temporal changes of Arctic sea ice coverage. The red line represents the annual average. **b,** Scatter plot of correlation between Arctic sea ice coverage and temperature. Each scatter point is the 5-day mean data (2018-2025), and the color represents the corresponding month.

## 4 Discussion

### 4.1 Uncertainties and limitations

The uncertainties of this study mainly stem from the inherent errors and parameterization biases of the multi-source datasets used, among which the uncertainties of the Dynamic World dataset and the ERA5 reanalysis dataset have the most significant impacts on the results. For the Dynamic World dataset, the accuracy of its nine LULC categories was verified using an expert-consensus confusion matrix, with the specific producer's accuracy and user's accuracy detailed in Table 4. The verification results show that the classification accuracy of most categories exceeds 70%, and the overall classification accuracy of the dataset reaches 73.8% (Brown et al., 2022), indicating that it has the basic accuracy required for conducting global-scale dynamic monitoring of land cover. However, uncertainties in this dataset still exist objectively, mainly reflected in the problem of category confusion in heterogeneous landscape areas. Misjudgments are prone to occur at the classification boundaries between low-coverage vegetation and bare ground, as well as between flooded vegetation and water bodies (Venter et al., 2022). Additionally, in high-latitude regions, classification errors are relatively higher due to the influence of cloud cover and light conditions, which may further lead to deviations in the calculation of key indicators such as forest coverage.

**Table 4.** Accuracy statistics from confusion matrix of Dynamic World (unit: %).

|  | Water | Forest | Grass | Flooded vegetation | Crops | Shrub | Built | Bare | Snow & Ice | Overall |
| --- | --- | --- | --- | --- | --- | --- | --- | --- | --- | --- |
| Producer's | 94.3 | 93.2 | 33.8 | 36.8 | 60.0 | 44.7 | 88.4 | 63.9 | 94.2 | 73.8 |

| | | | | | | | | | |
|---|---|---|---|---|---|---|---|---|---|
| accuracy | | | | | | | | | |
| User's accuracy | 90.6 | 70.2 | 30.1 | 65.2 | 88.9 | 54.7 | 86.7 | 60.8 | 71.2 |

The ERA5 sea ice cover dataset is essentially a reanalysis product that assimilates sea ice concentration (SIC) data from the EUMETSAT Ocean and Sea Ice Satellite Application Facility (OSI SAF). Consequently, its accuracy and inherent uncertainties largely derive from this foundational satellite-based dataset (Asadi et al., 2022). For dense ice zones of OSI SAF data, the mean absolute error (MAE) of SIC retrieval in both the Arctic and Antarctic is less than 5%, and the integrated global sea ice area error is constrained to ≤ 1% during summer months (Lavergne et al., 2019). Such overall accuracy provides a reliable basis for ERA5's global sea ice dynamic monitoring. Nevertheless, the inherent uncertainties of the OSI SAF dataset propagated to the ERA5 reanalysis product still cannot be ignored. In thin ice zones and marginal ice zones, the passive microwave retrieval algorithm of OSI SAF suffers from systematic underestimation of SIC. Fully covered thin sea ice is frequently misclassified as low-concentration ice, leading to an additional 3%–5% underestimation in the marginal ice zone. Importantly, these uncertainties primarily influence fine-scale and regional sea ice characterization, whereas the overall trends in global and polar sea ice cover dynamics captured by ERA5 remain robust. The magnitude of these biases falls within an acceptable range for conducting global-scale Earth system safety threshold analyses.

Beyond the dataset-specific uncertainties, this study is also subject to limitations stemming from methodological design and result interpretation. Regarding spatial scale trade-offs, the uniform 10 km resolution adopted (after resampling high-resolution input data) balances computational efficiency and global-scale analysis but smooths fine-scale heterogeneity (e.g., fragmented forests, narrow coastal sea ice fringes). This smoothing may underestimate indicator variability in heterogeneous regions, limiting the results' applicability to local or regional management. In addition, the driving mechanism analysis of cover change remains preliminary: while temperature was identified as a key driver of sea ice variation via correlation analysis, the relative contributions of other factors (e.g., anthropogenic emissions, ocean currents) were not quantified; for land cover changes, only qualitative attributions to deforestation or agricultural rhythms were provided, lacking a quantitative model to disentangle driver impacts.

## 4.2   Policy implications

The near-real-time global land-ocean cover dynamics monitoring framework developed in this study provides policymakers with a critical early-warning and assessment tool. The findings indicate that the global proportion of forested area relative to the original forest cover (averaging about 60%) is approaching the lower bound of the Earth system's safe limit (54%) and has shown a declining trend in recent years. Concurrently, Arctic sea ice coverage has already touched the upper safety threshold during certain months. This suggests that key ecosystems such as forests and the marine cryosphere are at risk of nearing their stability boundaries (Rockström et al., 2023). To achieve sustainable development within the "planetary boundaries", international and national governments need to establish a dynamic early-warning system based on the indicators provided by this study. Policymaking should integrate the Earth system safety perspective as a core principle, incorporating real-time data on changes in indicators like forest coverage and sea ice area into national and global-scale environmental assessment frameworks. Utilizing the data stream from this framework, timely ecological impact assessments and adjustments can be made for activities such as deforestation, agricultural expansion, and Arctic shipping route development, preventing single-sector policies from pushing key indicators beyond critical thresholds. Furthermore, decision-making should shift from passive damage response to proactive, threshold-based adaptive management, ensuring that human activities remain within ecological safety constraints.

The spatiotemporal patterns of global land-ocean cover provide a scientific basis for differentiated regional governance and climate adaptation strategies. The analysis shows significant disparities in the dominant patterns and seasonal rhythms of land cover types across different countries and regions, with Antarctic and Arctic sea ice variations exhibiting opposite phases. This calls for moving beyond a "one-size-fits-all" global governance model towards implementing refined, regionalized policies. For regions where forest coverage has already fallen below the safety threshold (e.g., around some urban agglomerations and agricultural development zones), policy priorities should shift towards ecological restoration and stringent land-use control (Yusof et al., 2023). For Arctic nations significantly impacted by sea ice changes, adaptive maritime management, shipping, and coastal community development plans should be formulated, and international attention should be directed to the potentially irreversible climate risks associated with the "near ice-free Arctic" tipping point. Additionally, this framework supports national-scale dynamic monitoring, which can assist countries in formulating differentiated strategies according to climatic, ecological, and socioeconomic contexts (He et al., 2025), balancing economic development and ecological protection to enhance long-term resilience under global warming.

## 5 Data availability

The global land-ocean cover dynamic dataset produced in this paper is available at https://doi.org/10.5281/zenodo.19228671 (Wang et al., 2026).

## 6 Conclusion

This study developed a near-real-time, coupled monitoring framework for global land-ocean cover dynamics by integrating multi-source remote sensing and reanalysis data. The work involved generating a long-term time series dataset (2018-2025) of major land cover types and sea ice coverage with a 5-day temporal resolution and near real-time update capability (lag < 1 month). Leveraging the GEE platform, a comprehensive processing workflow was established, enabling systematic spatiotemporal analysis at global and national scales, coupled with quantitative evaluations of key ecosystem safety thresholds and correlation with average temperature.

Our main findings reveal the intricate dynamics of the global land-ocean system. Spatially, distinct latitudinal gradients and regional differentiation patterns in land cover were observed, with forests dominating (27.0% of global land area) in tropical/subtropical regions. Temporally, the area proportions of vegetated cover types (e.g., forest, cropland) exhibited seasonal cycles peaking in the Northern Hemisphere summer, inversely correlated with non-vegetated types like snow and ice. Notably, long-term trend analysis indicates that forest, flooded vegetation, and cropland have declined in recent years, accompanied by expanded built-up areas and bare land, suggesting widespread conversion from ecological land to human-dominated and degraded surfaces. At the national scale, land cover composition and seasonal rhythms showed significant variation, reflecting divergent climatic conditions and human activity intensities. Furthermore, a pronounced anti-phase seasonal pattern was identified between Arctic and Antarctic sea ice coverage. Crucially, the safety threshold analysis indicates that the global forest cover indicator has been approaching the lower bound of its safe limit with a declining trend in recent years. Concurrently, Arctic sea ice coverage has intermittently touched its upper safe limit. Correlation analysis confirms temperature as a key driver of sea ice variation, showing a significant negative correlation ($R = -0.78$, $p < 0.001$), while thermodynamic evidence points to asymmetric rates of freezing and thawing processes.

The principal contributions of this research are threefold. First, it provides a novel, operational framework for the integrated near real-time monitoring of coupled land-ocean cover dynamics, addressing the gap in existing studies that often treat these components separately. Second, it delivers a high-timeliness dataset that captures the dynamic evolutionary processes of global cover types, supporting rapid assessment needs in the context of intensifying climate extremes. Third, it pioneers the integration of the Planetary Boundaries perspective into a dynamic monitoring framework. By quantifying the proximity of key indicators like forest and sea ice cover to their Earth system safety thresholds, this work transitions monitoring from mere observation to proactive risk assessment. Collectively, these findings provide essential data support and a scientific foundation for global climate change response, ecological safety evaluation, and the formulation of differentiated, adaptive sustainability policies.

**Supplement**

Figs. S1-3.

**Author contributions**

**L.W.:** Methodology, Software, Data Processing & Curation, Visualization, Writing - Original Draft; **T.L.:** Visualization, Writing - Review & Editing; **, X.D.:** Conceptualization, Writing - Review & Editing; **Z.L.:** Conceptualization, Supervision, Funding acquisition, Writing - Review & Editing.

**Competing interests**

The contact author has declared that none of the authors has any competing interests.

# Supplementary information

Near real-time monitoring of global land-ocean cover dynamics

**This file contains:**
Figs. S1-3.

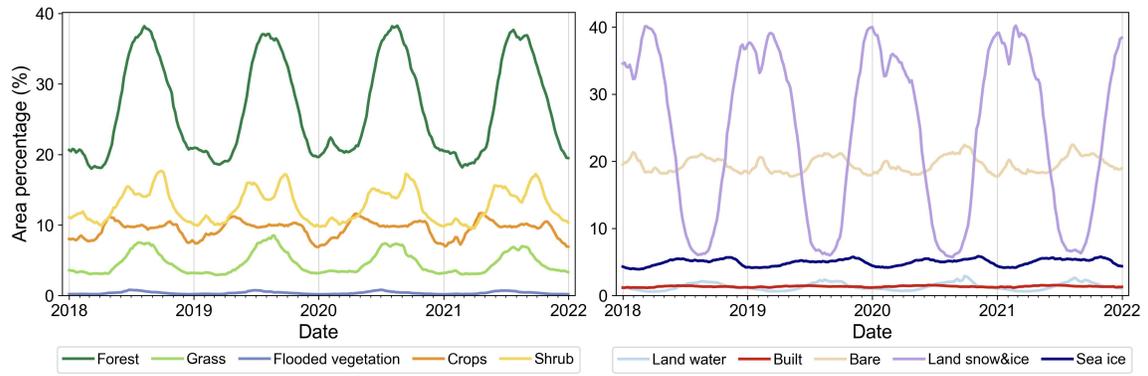

**Fig. S1.** The temporal change of the area percentage of each land-ocean type from 2018 to 2021.

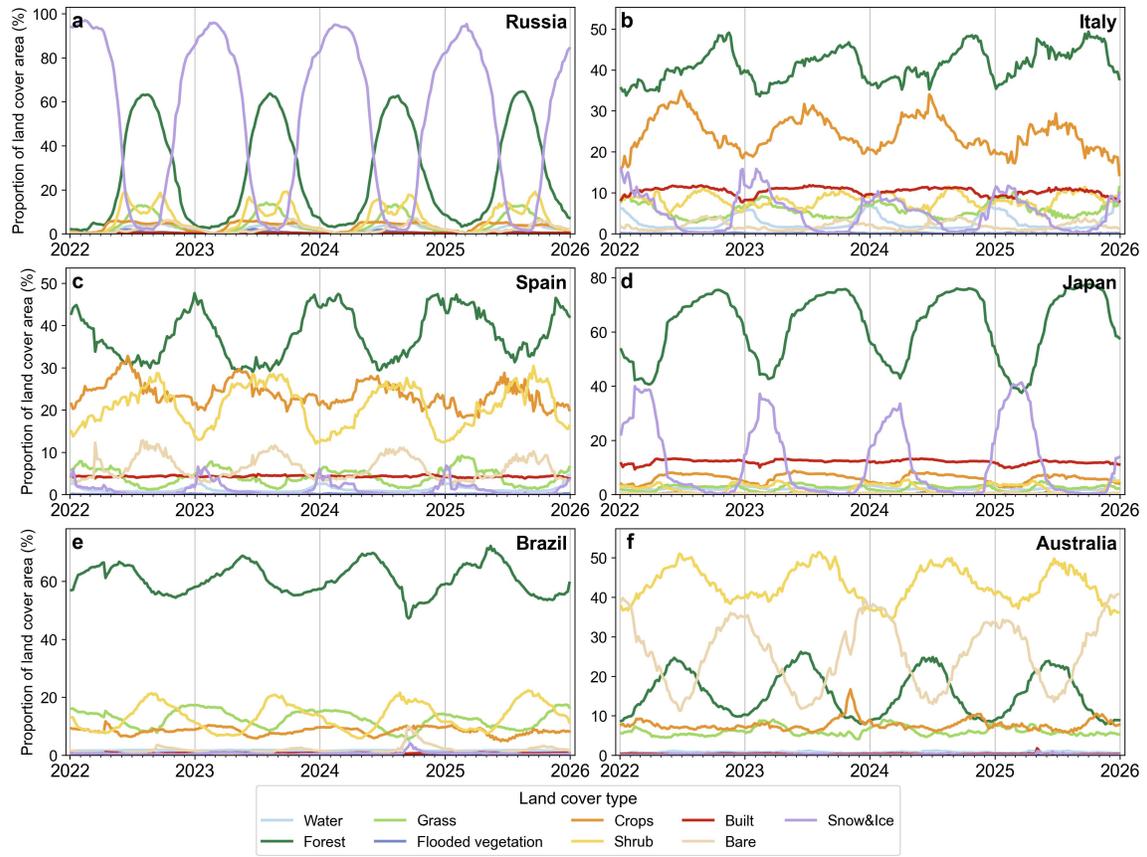

**Fig. S2. a-f,** Temporal changes of land cover in 6 major countries (Russia, Italy, Spain, Japan, Brazil and Australia).

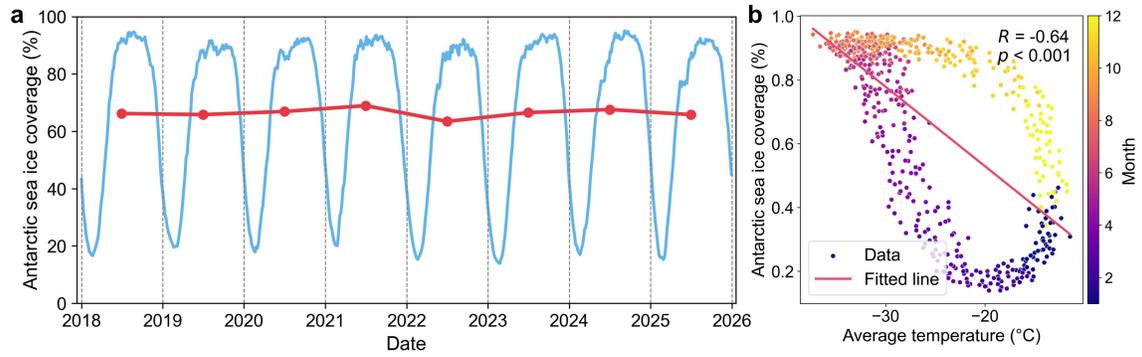

**Fig. S3. a,** Temporal changes of Antarctic sea ice coverage. The red line represents the annual average. **b,** Scatter plot of correlation between Antarctic sea ice coverage and temperature. Each scatter point is the 5-day mean data (2018-2025), and the color represents the corresponding month.